\documentclass[
	aip,
	jmp,
	amsmath,
	amssymb,
	reprint
]{revtex4-1}

\pdfoutput=1

\usepackage{graphicx}
\graphicspath{ {Figures/} }
\usepackage{textcomp}
\usepackage[per-mode = symbol, separate-uncertainty = true]{siunitx}
	\DeclareSIUnit\parsec{pc}
	\DeclareSIUnit\rad{\radian}
	\DeclareSIUnit{\rhz}{\ensuremath{\sqrt{\text{\hertz}}}}
\usepackage[breaklinks = true, colorlinks = true]{hyperref}
\usepackage[capitalize]{cleveref}

\begin{document}

\addtolength{\tabcolsep}{5pt}


\title{Small optic suspensions for Advanced LIGO input optics and other precision optical experiments}



\author{G. Ciani}
\email{ciani@phys.ufl.edu}

\author{M. A. Arain}
\altaffiliation[Current address: ]{Carl Zeiss Meditec, Dublin, CA-94568, USA}
\affiliation{University of Florida, Gainesville, FL-32611, USA}

\author{S. M. Aston}
\affiliation{LIGO Livingston Observatory, Livingston, LA-70754, USA}

\author{D. Feldbaum}

\author{P. Fulda}

\author{J. Gleason}

\author{M. Heintze}

\author{R. M. Martin}

\author{C. L. Mueller}

\author{D. M. Nanda Kumar}
\affiliation{University of Florida, Gainesville, FL-32611, USA}

\author{A. Pele}
\affiliation{LIGO Livingston Observatory, Livingston, LA-70754, USA}

\author{D. H. Reitze}
\altaffiliation[Current address: ]{LIGO Laboratory, California Institute of Technology, Pasadena, 
	CA-91125, USA}

\author{P. Sainathan}

\author{D. B. Tanner}

\author{L. F. Williams}

\author{G. Mueller}
\affiliation{University of Florida, Gainesville, FL-32611, USA}

%

\date{\today}

\begin{abstract}
We report on the design and performance of small optic suspensions developed to suppress seismic motion of out-of-cavity optics in the Input Optics subsystem of the Advanced LIGO interferometric gravitational wave detector. These compact single stage suspensions provide isolation in all six degrees of freedom of the optic, local sensing and actuation in three of them, and passive damping for the other three.
\end{abstract}


\maketitle 

\section{Introduction}\label{sec_Intro}
On September 14th, 2015, at the beginning of their first observing run, the Advanced LIGO Gravitational Wave detectors made the first direct detection of gravitational waves \cite{GW150914}. For about 4 months, although not yet at full sensitivity, the two instruments routinely operated with a range between \num{70} and \SI{80}{\mega\parsec} for a reference NS-NS binary system, observing a volume more than \num{50} times larger than their predecessors \cite{aLIGOSensitivity}.
Critical to this success has been the performance of the Input Optics (IO) subsystem \cite{InputOptics}, designed, built, installed and tested by the LIGO group at the University of Florida; the subsystem is charged with delivering a stable and well-shaped beam to the main interferometer across the whole range of possible operating input powers, up to \SI{180}{\W}. The in-vacuum portion of the IO subsystem employs \SI{75}{\mm} diameter optics to steer and mode-match the laser beam from the input mode cleaner into the power recycled interferometer; these out-of-cavity optics are suspended by small, single stage vacuum compatible suspensions called HAM auxiliary suspensions (HAUX) to isolate them from residual vibration of the optical table and to allow for pointing and local damping.

Although developed in the context of Advanced LIGO, the HAUX suspensions can find application in a broader range of lab-scale optical experiments. They provide isolation in all degrees of freedom, local sensing and actuation, and active and passive damping, while employing a compact, lightweight mechanical design with a number of expedients to simplify operation and maintenance of the suspension and the installed optic.

This paper describes the requirements, design, and performance of the HAUX. \Cref{sec_ReqConcDes} lists the performance requirements and desired operating characteristics, and explains how they have driven the top level design choices. \Cref{sec_MecDesign} describes the mechanical setup in greater detail, and the design expedients put in place to make assembly and maintenance of the suspension more convenient. \Cref{sec_Performance} presents data from the main performance tests performed on typical HAUX production units; finally, \cref{sec:conclusions} concludes this manuscript by summarizing the HAUX main strengths.

Detailed documentation regarding the requirements, final design and testing of the HAUX can be accessed from the LIGO Document Control Center\cite{DCC} under entry number LIGO-E1600169\cite{DCCHAUXref}.

\section{Requirements and conceptual design}\label{sec_ReqConcDes}
\begin{figure}[t]
	\includegraphics[width=\linewidth]{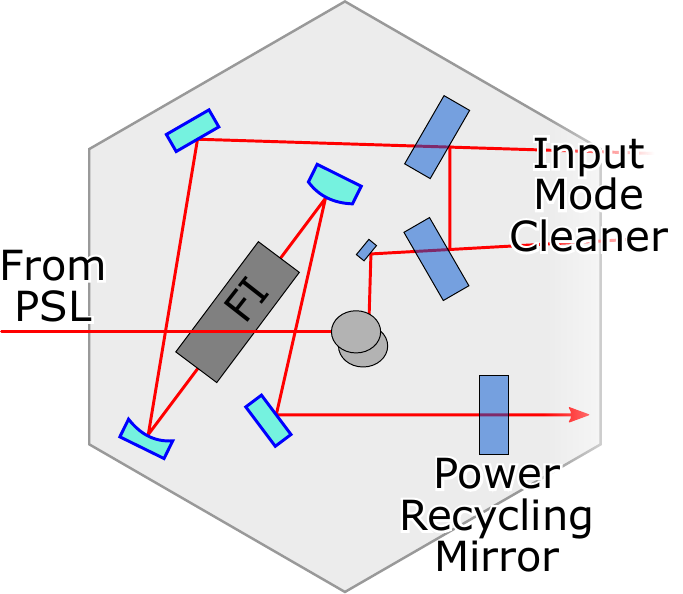}
	\caption{\label{fig_HAM2}
		(Color online) A simplified representation of the HAM2 table which hosts the majority of the in-vacuum input optics. The laser beam coming from the the pre-stabilized laser (PLS), on the left in the picture, propagates over the Faraday isolator (FI) and is lowered to table-height via the periscope in the center of the table. A small fixed optic steers it into the triangular input mode cleaner (IMC), of which only two optics are shown here. The mirror installed in the four HAM Auxiliary Suspension, depicted in light blue on the left portion of the HAM2 table, route the beam, filtered and stabilized by the IMC, through the FI and towards the power recycling mirror. Two of these mirrors are curved and provide mode matching to the main interferometer.
		}
\end{figure}

As shown in \cref{fig_HAM2}, the four optics suspended by the HAUX are all located on a single Advanced LIGO seismic isolation table, after the input mode cleaner (IMC) and before injection into the power-recycling cavity (PRC) of the main interferometer. The conceptual structure of the HAUX suspension chain has been based on the need of both precisely controlling the alignment of the input beam with respect to the main interferometer, and preserving the noise performance of the Input Optic Subsystem. In particular, the frequency and pointing noise introduced by the HAUX is generally required to be at least a factor \num{10} below the noise at the output of the IMC.

Noise in the displacement \textit{x} along the optical axis of a reflecting optic causes a variation of the total optical path and appears as frequency noise
$S^{1/2}_\nu = \frac{4\pi}{\lambda} f S^{\frac{1}{2}}_x$
in the beam downstream of the optic. In addition, rotational noise of the optic around the horizontal or vertical axis orthogonal to the optical axis (referred to as \textit{pitch} and \textit{yaw}, respectively) directly couples into beam pointing noise
$S^{1/2}_\theta = 2 S^{\frac{1}{2}}_{pitch,yaw}$
 (for a flat mirror).

Based on IMC requirements \cite{IObeamJitter}, assuming that the four mirrors experience uncorrelated noise and conservatively neglecting the effect of non-normal angle of incidence, for a single suspension this translates into a requirement on the residual displacement noise of

\begin{equation}
	\begin{aligned}
		S^{1/2}_x &\leq 2 \cdot 10^{-11} \si[per-mode=fraction]{\m\per\rhz} \, @ \, \SI{10}{\Hz} \\
		S^{1/2}_x &\leq 4 \cdot 10^{-14} \si[per-mode=fraction]{\m\per\rhz} \, @ \, \SI{100}{\Hz} \\
	\end{aligned}
\end{equation}

and on the residual rotational noise expressed as

\begin{equation}
	S^{1/2}_\alpha \leq 6 \cdot 10^{-13} \sqrt{1+\left(\frac{\SI{100}{\Hz}}{f}\right)^4} \si[per-mode=fraction]{\radian\per\rhz}
\end{equation}

These requirements are valid above \SI{10}{Hz}, which is the lower limit of the Advanced LIGO measurement band.

In general, one degree of freedom of a single stage mechanical suspension can be modeled (under ideal conditions) as a harmonic oscillator with a natural resonant frequency $f_0$. For frequencies $f$ above resonance, the displacement response of the suspended mass to external forces quickly approaches that of a free mass, decaying as $1/\left(m \left(f_0^2-f^2\right)\right)$.
The response to motion of the suspension point shows a similar decay with frequency and goes as $f_0^2/(f_0^2-f^2)$. The level of isolation at a given frequency can thus be controlled by an appropriate choice of $f_0$.

\begin{figure}
	\includegraphics[width=0.8\linewidth]{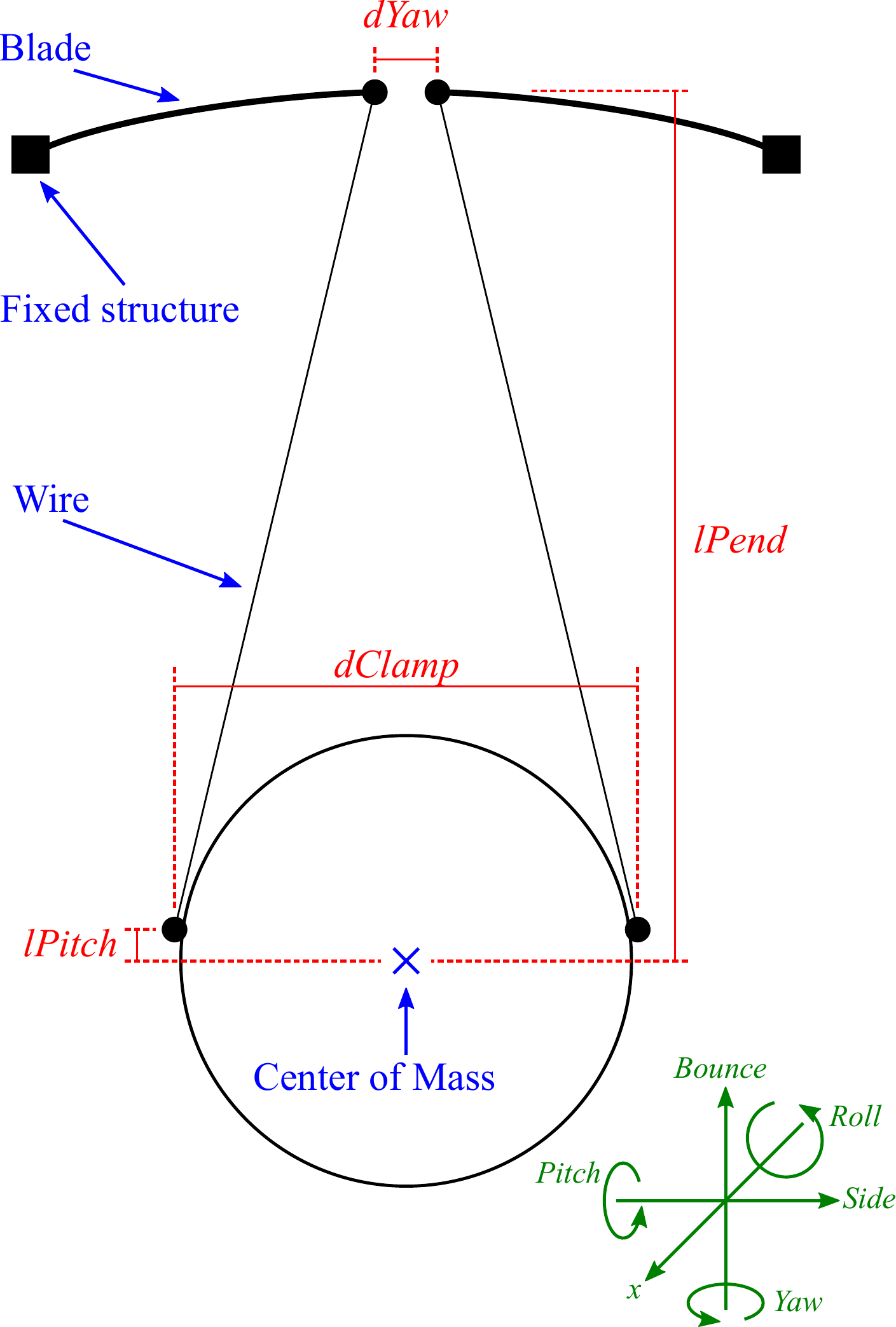}
	\caption{\label{fig_SuspChain}
		(Color online) A schematic representation of the HAUX suspension chain. The main functional elements are noted in blue. The six main degrees of freedom of the optic are identified in green in the bottom right corner. In red are shown the relevant dimensions that set the resonant frequency for \textit{x}, \textit{pitch} and \textit{yaw}.}
\end{figure}

Considering the expected performance of the Advanced LIGO seismic isolation stack on which the HAUX are mounted\cite{SEI2015}, and even accounting for a safety factor, the HAUX requirements can be easily fulfilled by providing a single stage of isolation for \textit{x}, \textit{pitch} and \textit{yaw}, as long as the resonance frequency in each degree of freedom is kept below a few Hz. 
\Cref{fig_SuspChain} shows a schematic representation of the configuration chosen for the HAUX suspensions, together with the geometrical parameters relevant to their performance.
The optic is suspended from two wires, one on either side, which come close together at the upper suspension point; to a first approximation the resonant frequency of \textit{x} is controlled by the length of the suspension wires (\textit{lPend}), that of \textit{yaw} is controlled by the horizontal separations of the upper (dYaw) and lower (\textit{dClamp}) wire attachment points and that of \textit{pitch} is controlled by the height of the lower attachment points above the optic's center of mass (\textit{lPitch}).

Below the Advanced LIGO measurement band, from \SI{0.1} to \SI{10}{Hz}, the overall motion of the optic in \textit{pitch} and \textit{yaw} must be limited to \SI{1}{\micro\rad} RMS for the alignment sensing and control subsystem to be able to maintain the interferometer close to its ideal working point. This requires the amplitude of motion at the resonant frequencies to be limited, which is accomplished via active damping using a set of four sensor-actuators described in \cref{sec_MecDesign}. 

An additional requirement for the HAUX is that the remaining three degrees of freedom of the optic must each have at least a single stage of isolation from the motion of the table, with a resonant frequency below \SI{10}{\Hz}. This requirement is met by hanging the two suspension wires from two blade springs: the common or differential motion of the blade's tip allow for the optic to rotate around the optical axis (\textit{roll}), swing orthogonal to it (\textit{side}) or move along the vertical axis (\textit{bounce}). Again, adjusting the geometrical and mechanical parameters allows controlling the resonance of these three degrees of freedom.

While it is clear that this geometry arrangement provides all the physical parameters needed to adjust the resonances to desired values, calculating the final transfer functions is not straightforward; the intuitive picture of independent harmonic oscillators acting along the different degrees of freedom is useful to understand how the various resonant frequencies can be controlled, but is not accurate; in reality, \textit{x} and \textit{pitch}, as well as \textit{roll} and \textit{side}, are degrees of freedom of double oscillator systems, and each pair combine to form two normal modes. In addition, the longitudinal and bending stiffness of the wires have a non-negligible effect on the resonances of some of the modes\cite{CalumPhDTh}. To calculate the values of the geometrical parameters needed to obtain the desired resonant frequency for each mode we used a semi-analytical model implemented as a Mathematica package\cite{BartonMathematicaModels}, which accounts for all these effects. We also used the finite element analysis software COMSOL Multiphysics to model the blade springs and to obtain the desired vertical spring constant. The value of the parameters and the modeled resonant frequencies are reported in \cref{tab_HAUXparam,tab_HAUXfreq}. Note that, since precise matching of pre-determined resonance frequency values was not a requirement, no mechanism has been incorporated in the HAUX design to compensate for machining and assembly tolerances and to fine-tune the resonance frequency values.

\begin{table}
	\caption{\label{tab_HAUXparam}
		values of geometrical parameters used in the final design of the HAUX (top table).}
	\begin{tabular}{l  r}
		\hline
		\textit{Parameter} 	& Design value (mm) \\ \hline 
		\textit{dYaw} 		& 15.7 \\ 
		\textit{dClamp} 	& 100.3 \\ 
		\textit{lPitch} 	& 1.0 \\ 
		\textit{lPend} 		& 259.3 \\ \hline
	\end{tabular}
\end{table}
\begin{table}
	\caption{\label{tab_HAUXfreq}
		modeled and measured resonant frequencies of the six degrees of freedom of the optic; modeled results are calculated using exact nominal values from \cref{tab_HAUXparam}. Measured values reported here, which have a $1\sigma$ confidence interval equal to the last significant digit, refer to the final prototype; production units exhibits very similar values. Except for Yaw, the measured resonant frequencies are systematically lower than modeled; this discrepancy is further discussed in \cref{sec_Performance:resonanceAndTF}.}
	\begin{tabular}{l  r r}
		\hline
		Mode & Modeled f$_0$ (Hz) & Measured f$_0$ (Hz) \\ \hline 
		x/pitch 1 	& 0.98	& 0.95\\ 
		x/pitch 2 	& 1.12	& 1.04\\ 
		yaw 		& 0.76	& 0.80\\ 
		bounce 		& 7.19	& 6.14\\ 
		side/roll 1 & 1.00	& 1.00\\ 
		side/roll 2 & 10.63	& 8.97\\ \hline
	\end{tabular}
\end{table}

\section{Mechanical design}\label{sec_MecDesign}
The final HAUX mechanical design, in addition to accommodating the suspensions chain described in the the previous section, had to satisfy a number of functional demands:
\begin{itemize}
	\item being vacuum compatible at the particularly stringent level required for Advanced LIGO, which imposes restrictions not only on the total outgassing, but also on the molecular species being outgassed;
	\item being able to accommodate \SI{75}{\mm} diameter mirrors with a clear aperture of at least \SI{10}{\mm}, for horizontal angles of incidence up to \SI{55}{\degree};
	\item providing active control of the optics in \textit{x}, \textit{pitch} and \textit{yaw}, and passive damping for all other degrees of freedom;
	\item being equipped with safety stoppers to protect the optic in case of unexpected shaking or a wire failure, and to allow the optic to be clamped in place when needed;
	\item allowing for fine tuning the optic working position in \textit{pitch} to within \SI{1}{\milli\radian}, so as to mechanically remove any large offset from the active control system;
	\item having the lowest structural resonance above about \SI{150}{\Hz}, to avoid interfering with the seismic-isolation table's active control system;
	\item last but not least, being as compact and simple as possible.
\end{itemize}

\begin{figure}
	\centering
		\includegraphics[width=0.65\linewidth]{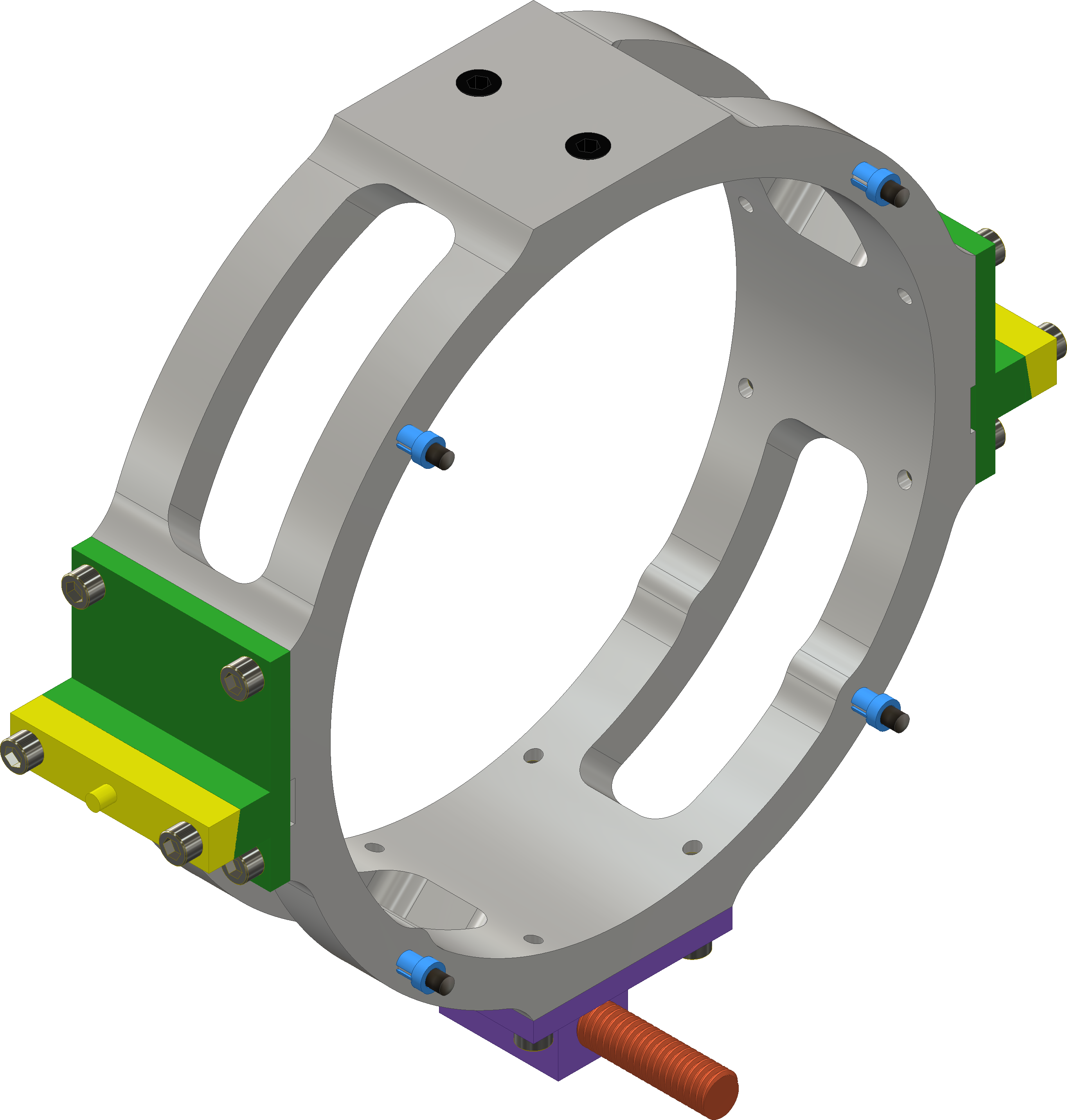}
	\quad\quad\quad\quad\quad
		\includegraphics[width=0.65\linewidth]{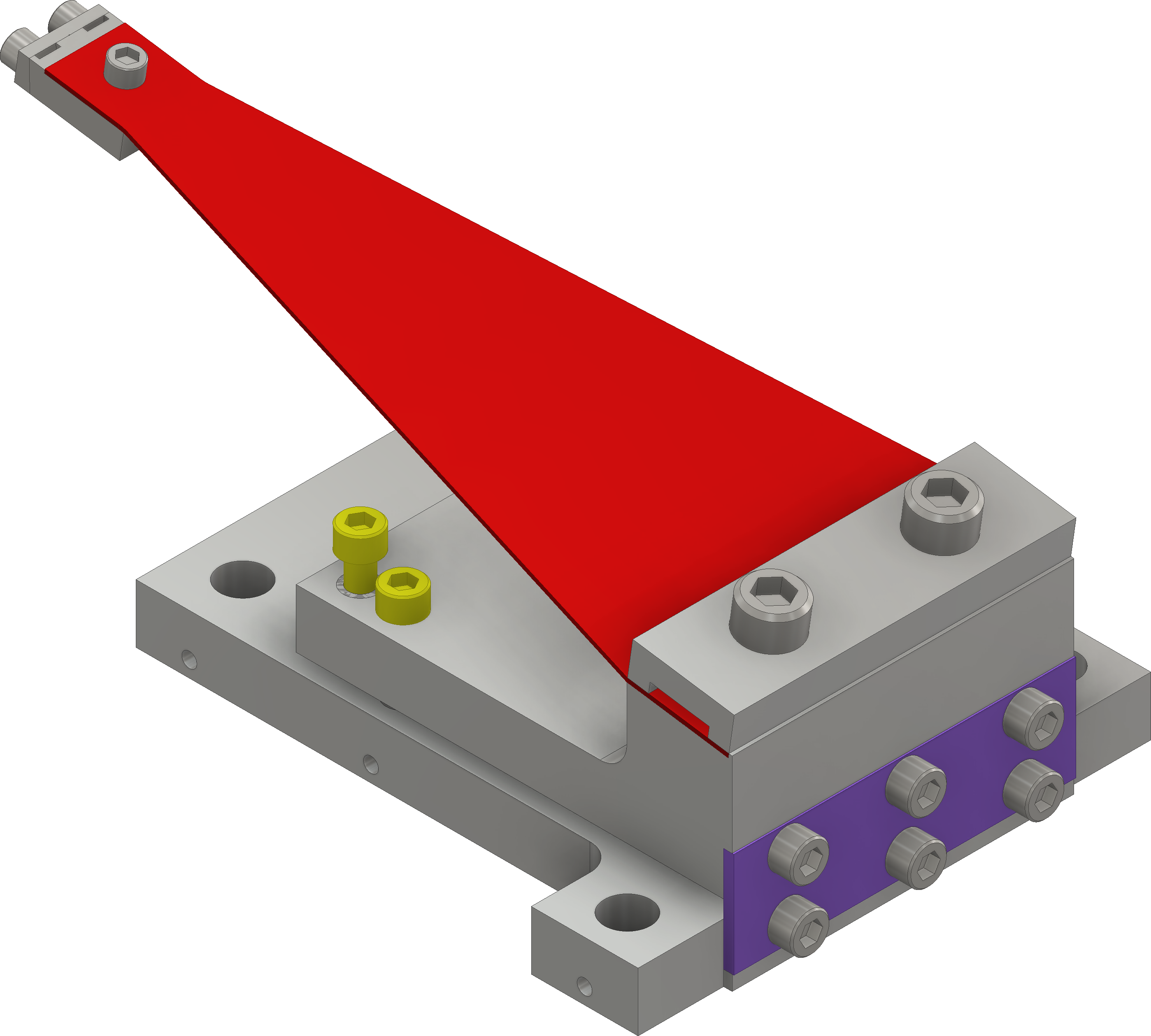}
	\caption{\label{fig_OpticHolderAndBlades}
		(Color online) Top: the aluminum optic holder used in the HAUX, seen from the back of the optic and represented in colors for clarity. Note the wire clamps (green and yellow) on the sides, the balancing rod (orange) at the bottom, and the magnets (black) together with their small steel standoffs (light blue). Also note the optic locking screws at the very top (black) and the four raised contact points at \SI{+-45}{\degree} from the bottom, aligned with the position of the bottom magnets. Bottom: a blade spring assembly. The blade spring itself is represented in red; in purple is the steel flexture plate, and in yellow the push-pull screws.}
\end{figure}

Assembly and maintenance of the suspensions and suspended optics is made more convenient by installing the \SI{75}{\mm} diameter, \SI{25}{\mm} thick optic in a lightweight aluminum holder, shown in \cref{fig_OpticHolderAndBlades} (top). In this way, attachment clamps and actuation magnets can be attached to the holder rather than directly glued to the optic, allowing for easy replacement of the mirror with minimal disassembly of the suspension.
The optic is held in position by a pair of PEEK set screws that push it from the top against four raised contact points distributed on the front and back of the holder, at \SI{+-45}{\degree} from the bottom.
A wire clamp provided with an array of pre-machined vertical grooves is attached on either sides of the aluminum holder. The vertical grooves help fix the position of the wire and avoid over-squeezing; the array is necessary to be able to accommodate for the tolerance in the physical dimension of the different optics by selecting the suspension point closer to the center of mass.
The optic holder has a fine-threaded horizontal through hole at the bottom that allows for a copper rod to be screwed back and forth to move the center of mass of the assembly by small amounts and adjust the static \textit{pitch} of the optic.

\begin{figure}
	\includegraphics[]{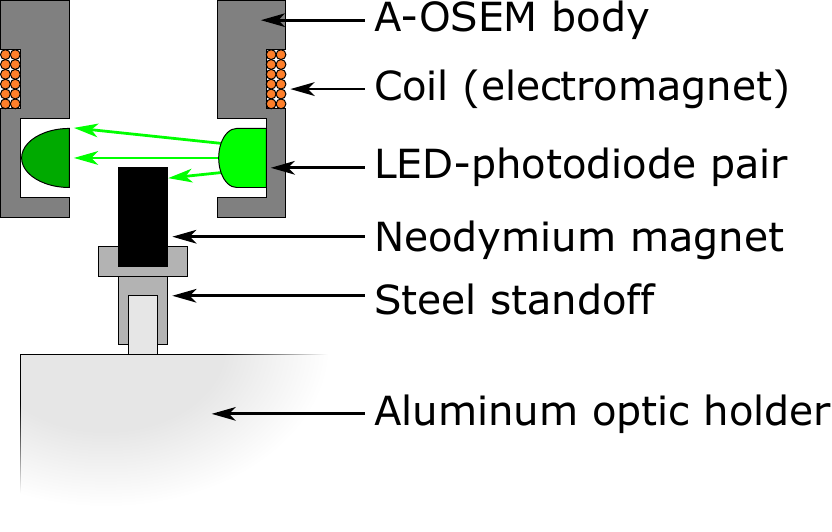}
	\caption{\label{fig_aosem}
		(Color online) A cross-section sketch of an AOSEM sensor/actuator. The body of the AOSEM is a bored cylinder that supports a coil on the outside and a LED-photodiode pair facing each other on the inside. The magnet attached to the optic holder is interposed between the LED and photodiode, partially shading the latter. As the optic moves, the amount of light reaching the photodiode varies and a readout can be obtained. Running a current through the coil creates a magnetic field that pushes or pulls on the magnet, thus actuating the optic.}
\end{figure}

Four magnets are magnetically attached to four steel sleeves, which are press-fitted over matching posts arranged in a \SI{58.2}{\mm} square pattern on the back of the holder. The magnets work together with the sensing/actuation units called AOSEMs \cite{OSEMsCQG}. As shown in \cref{fig_aosem}, an AOSEM is a combination of an electromagnetic actuator and a shadow sensor, in turn comprised of a LED and a photodetector. The AOSEM is attached to the structure of the suspension in such a way that the magnet is coaxial with the coil and partially shields the photodetector from the light coming from the LED. As the magnet moves back and forth, the amount of light reaching the photodetector changes and a position readout can be obtained. At the same time, a current can be run through the coil, creating a magnetic field and a force on the magnet.
By suitable linear combinations of the readout of the four AOSEMs, signals for \textit{x}, \textit{pitch} and \textit{yaw} can be obtained (three AOSEMs would be sufficient, but a fourth one has been introduced for redundancy, calibration and symmetry purposes).

Two \SI{250}{\mm} long, \SI{150}{\um} diameter steel music wires run from the bottom clamps to smaller, single-groove clamps at the tips of the spring blades, shown in \cref{fig_OpticHolderAndBlades} (bottom).
The width profile of the monolithic, \SI{0.5}{\mm} thick stainless steel blades is comprised of three sections: a \SI{9.5}{\mm} wide, \SI{9}{\mm} long flat tip attached to the wire clamp, a \SI{76.8}{\mm} long tapered section that is free to flex and a \SI{40.6}{\mm} wide, \SI{12}{\mm} long section clamped to an adjustable blade support. The variable width of the tapered section allows for the stress to be equally distributed along the blade when the tip is loaded\cite{DCCblades}.
The support has a flexible joint and a system of push-pull screws that provide fine control of the departure angle of the blade, so that the working point of the tip can be made to be horizontal under load and, for small angles, oscillate only in the vertical direction. According to an analytical study and a finite-element model, the resulting spring constant is \SI{380}{\N\per\m} and the maximum stress under load is approximately \SI{85}{\mega\Pa}, less than 50\% of the yield strength of stainless steel.

\begin{figure*}
	\centering
	\includegraphics[width=\linewidth]{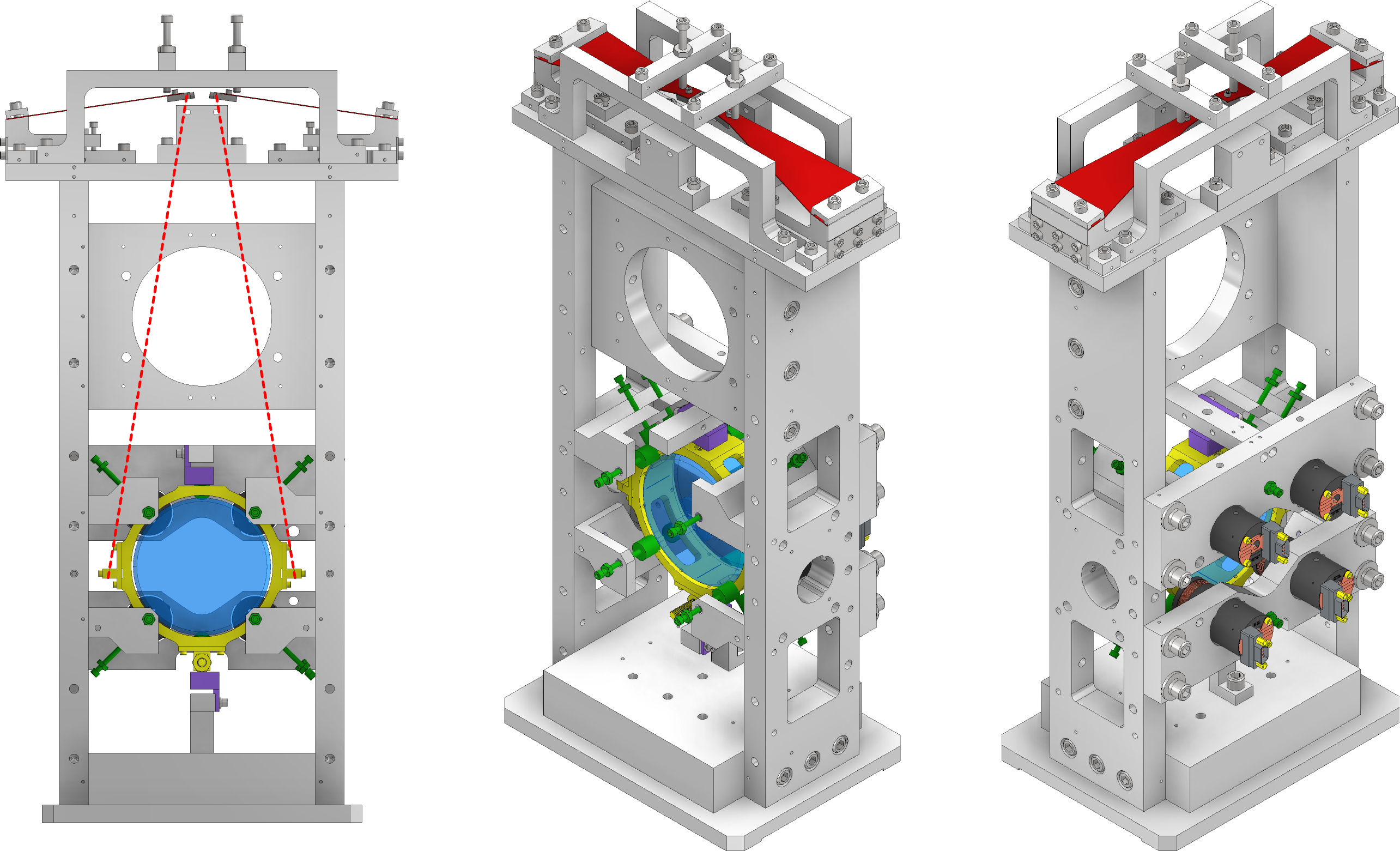}
	\caption{\label{fig_HAXU3D}
		(Color online) front and isometric views of a 3D model of the suspension, with relevant components highlighted in color: optic in blue; optic holder in yellow; locking screws in green; damping magnet casings in purple; AOSEMs in dark grey; blade springs in red. The wires are not shown in the 3D model, and the blades are represented unloaded (straight); the approximate placement of the wires is shown by the dashed red line in the front view.}
\end{figure*}

\Cref{fig_HAXU3D} shows a 3D model of an assembled HAUX suspension. The main structure, made of aluminum to reduce weight, fits in an envelope of dimensions \SI{127 x 217 x 441}{\mm} ($D \times W \times H$) and weighs approximately \SI{6}{\kg}. It consists of a base, two side walls, two horizontal bars each supporting two AOSEMs, a stiffening slab connecting the two walls and a top slab to which the blade supports are attached (using slotted holes to allow for precise adjustment during assembly).
From the structures surrounding the optic, a set of 14 soft-tip screws protrude towards the aluminum holder and serve the purpose of safety stop and clamping devices in case of need.
Two pairs of neodymium magnets are mounted in aluminum casings directly above and below the optic holder, and provide passive damping via eddy currents. The magnets are anti-parallel, so as to minimize the field far from the suspension, and their distance from the holder can be adjusted from \num{0} to about \SI{5}{\mm} to obtain the desired level of damping. It should be noted that, although the damping is intended to mainly affect the degrees of freedom not actively controlled by the AOSEMs, this arrangement of magnets has some damping effect also on \textit{x}, \textit{pitch} and, much less, \textit{yaw}.

\section{Performance}\label{sec_Performance}
The final HAUX prototype, as well as the production units, have been tested for both compliance with the requirements and quality of manufacturing \cite{DCCHAUXmodTest,DCCL1HAUXtest,DCCH1HAUXtest}. The following subsections highlight the most important tests; when there are no significant differences between the various units, in the interest of clarity and space only a representative subset of results is presented.

Absolute calibration of most of these measurement involves independent calibration of a variety of software and hardware interfaces which are part of the Advanced LIGO control infrastructure, and not directly related to the HAUX design or their performance. Such calibration was often not available at the time the measurement were taken. For this reason, data are presented either in arbitrary units, or with a nominal calibration, depending on the case. Once the suspensions have been installed in the interferometer and the final components were available, the calibration of the entire chain, from software control interface to mirror response, has been verified to within 15\% from the nominal value \cite{LLOIM3cal}.

\subsection{Pitch and yaw pointing range}\label{sec_Performance:range}
\begin{figure}
	\includegraphics[width=\linewidth]{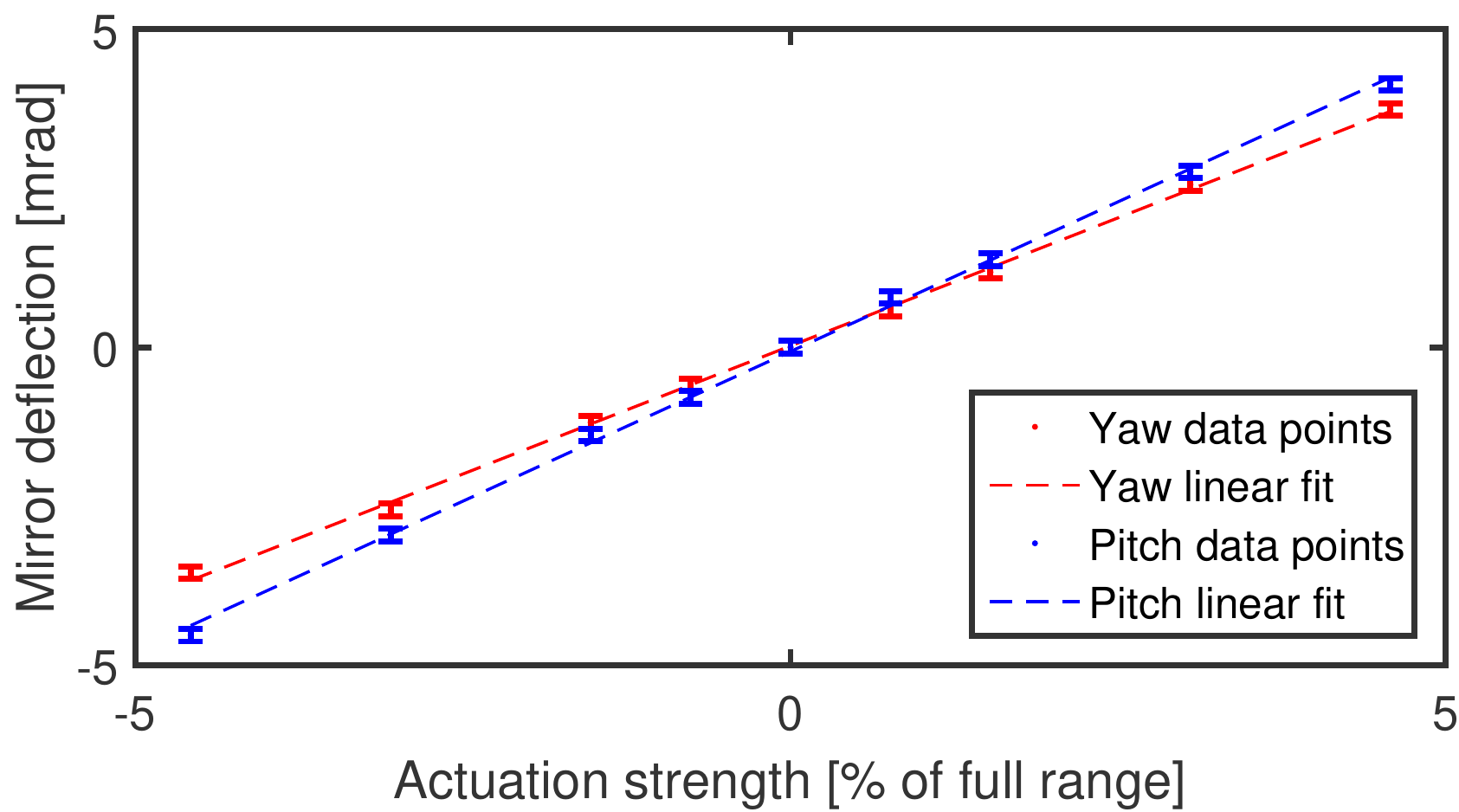}
	\caption{\label{fig_OSEMact}
		(Color online) Plot of the measured mirror deflection angle as a function of commanded actuation (arbitrary units). The response is linear over the entire actuation range of \SI{+-5}{\milli\radian} (\SI{+-10}{\milli\radian} of beam deflection). The actuation is expressed in arbitrary units because calibration was not available for the preliminary electronic chain used at the time of this measurement.}
\end{figure}

The mechanical limit to \textit{pitch} and \textit{yaw} rotation of the optic is set by the position of the safety stops, which can be adjusted to accommodate a range well beyond \SI{10}{\milli\radian}. The practical limit is then set by the available current to the coils, by the force per unit current that the AOSEMs can exert, \SI{16e-3}{\N\per\A} with the magnets in use, and by the rotational stiffness of pitch and yaw, which with the design values for \textit{dYaw}, \textit{dClamp} and \textit{lPitch} is of the order of \SI{5e-3}{\N\m\per\rad} for both degrees of freedom. The target dynamic range of \SI{+-10}{\milli\rad} can thus by obtained with a current of \SI{35}{\milli\A}, or the maximum range can be reduced improving noise and resolution, depending on the design of the current driver.
The linearity of the actuation in the range of interest is also important. \Cref{fig_OSEMact} shows the rotation of the optic measured using an optical lever for one of the production units, for various values of the commanded actuation.

\subsection{Resonant frequencies and transfer functions}\label{sec_Performance:resonanceAndTF}
The measured resonant frequencies for all six degrees of freedom are reported in \cref{tab_HAUXfreq} alongside the modeled values. These measurements were performed using the final prototype suspension built; however, the other eight production units assembled and installed in the Advanced LIGO detectors exhibit very similar values.
It is apparent that there is a systematic tendency of the measured frequencies to be lower than the modeled ones. For the degrees of freedom dominated by the blades spring motion this is expected, as the blades had been independently tested before installation, and found to be softer than originally modeled (probably because of slightly different physical properties of the material used). For the other degrees of freedom the origin of the discrepancy has not been identified. A study of the suspension computer model has shown that a number of realistic machining and assembly tolerances, as well as material property variations, can combine in several ways to explain the observed values. However, this has not been further investigated, since it was of no particular interest in the context of Advanced LIGO where lower resonant frequencies are actually an advantage.

\begin{figure}
	\centering
	\includegraphics[width=\linewidth]{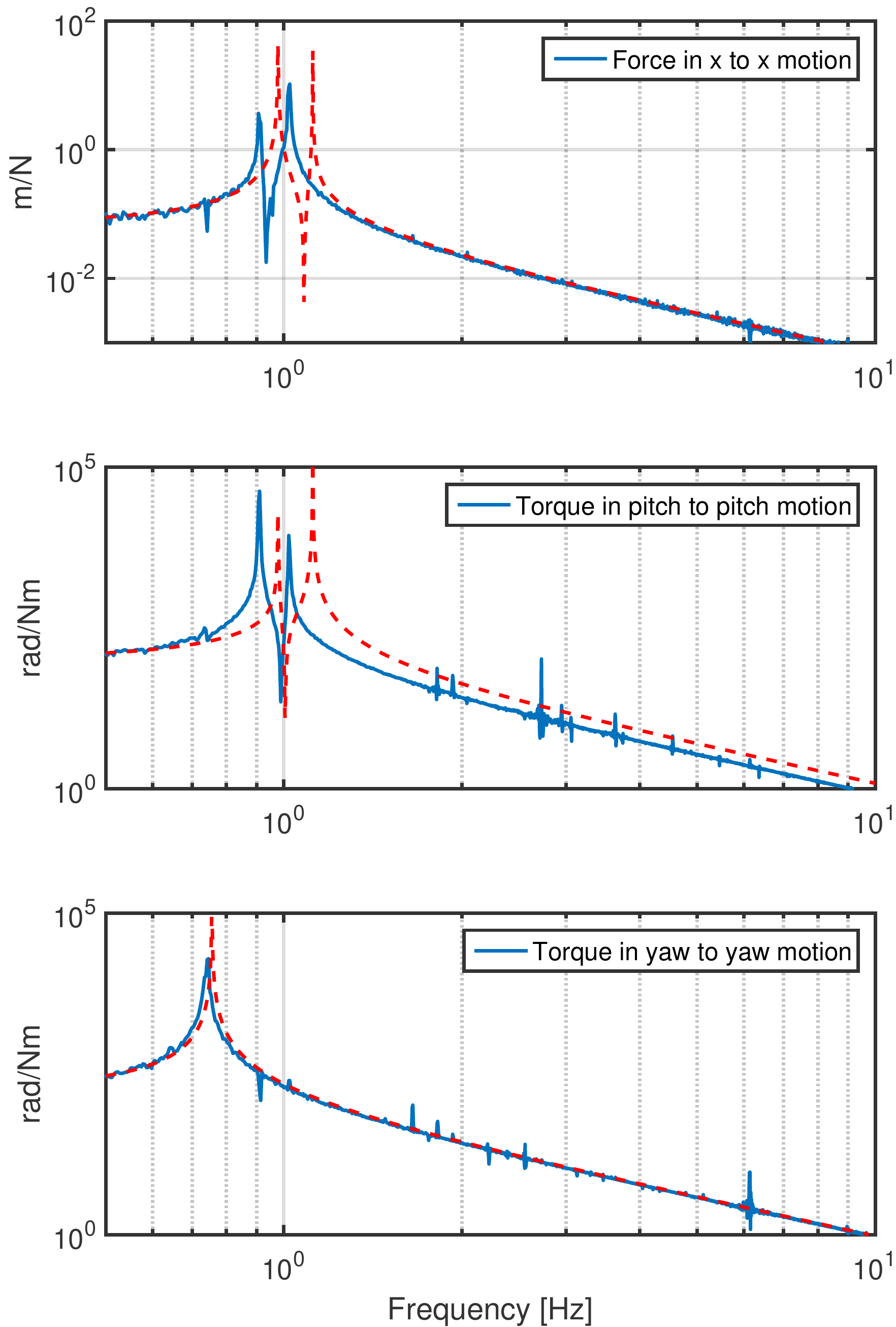}
	\caption{\label{fig_TFs}
		(Color online) Example transfer function from force (torque) to displacement (rotation) for one of the production units, LHO\_IM3.}
\end{figure}

\begin{figure}
	\includegraphics[width=\linewidth]{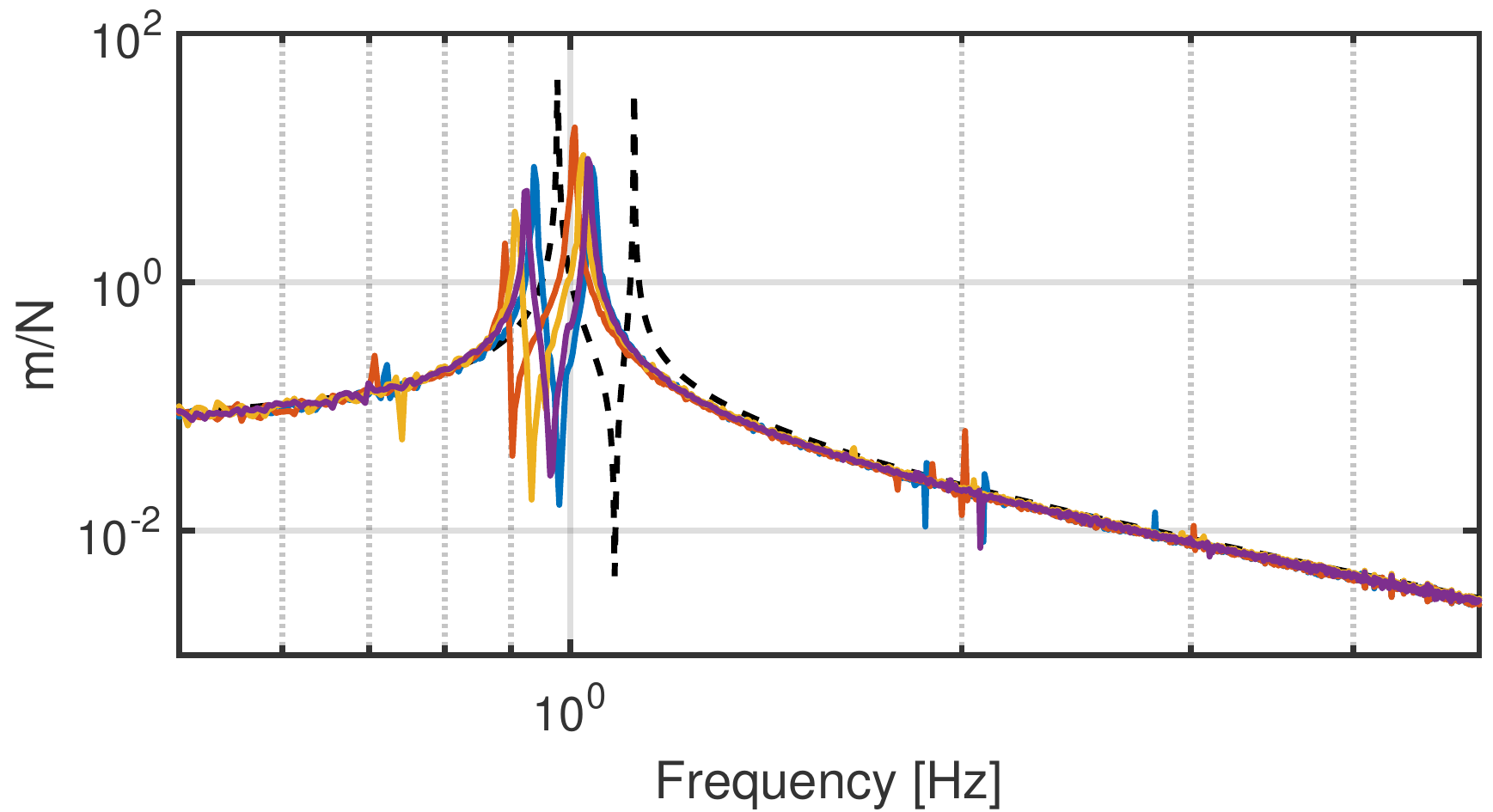}
	\caption{\label{fig_TFs_comparison}
		(Color online) Solid lines: transfer functions from force to motion in \textit{x}, for the four different units installed at the LIGO Handford Observatory. Dashed line: model.}
\end{figure}

With the exception of the decrease in the resonant frequencies discussed above, the measured transfer functions agree very well with the model. As an example, \cref{fig_TFs} shows a subset of transfer functions from force (or torque) to displacement (or rotation) of the optic for one particular production unit labeled H1-IM3. Again, all the assembled and tested units show comparable results, as exemplified in \cref{fig_TFs_comparison}, which shows the force to motion in the \textit{x} degree of freedom transfer function, measured for 4 different units at the LIGO Hanford Observatory. Given that the HAUXs do not incorporate any mechanism to fine-tune the resonances of the as-built units, the agreement is very satisfactory, and compatible with expected machining and assembly tolerances.

It should be pointed out that the main goal of the suspensions is that of isolating the motion of the optic from that of the suspension point, making the transfer functions shown here not be the ones we are most directly interested in. However, they can be measured more easily and accurately than transfer functions from motion of the suspension to motion of the optics, and being dependent on the same parameters, they provide an equally valid verification of the suspension performance and agreement with the model. 

\subsection{Active and passive damping}
\begin{figure}
	\centering
	\includegraphics[width=\linewidth]{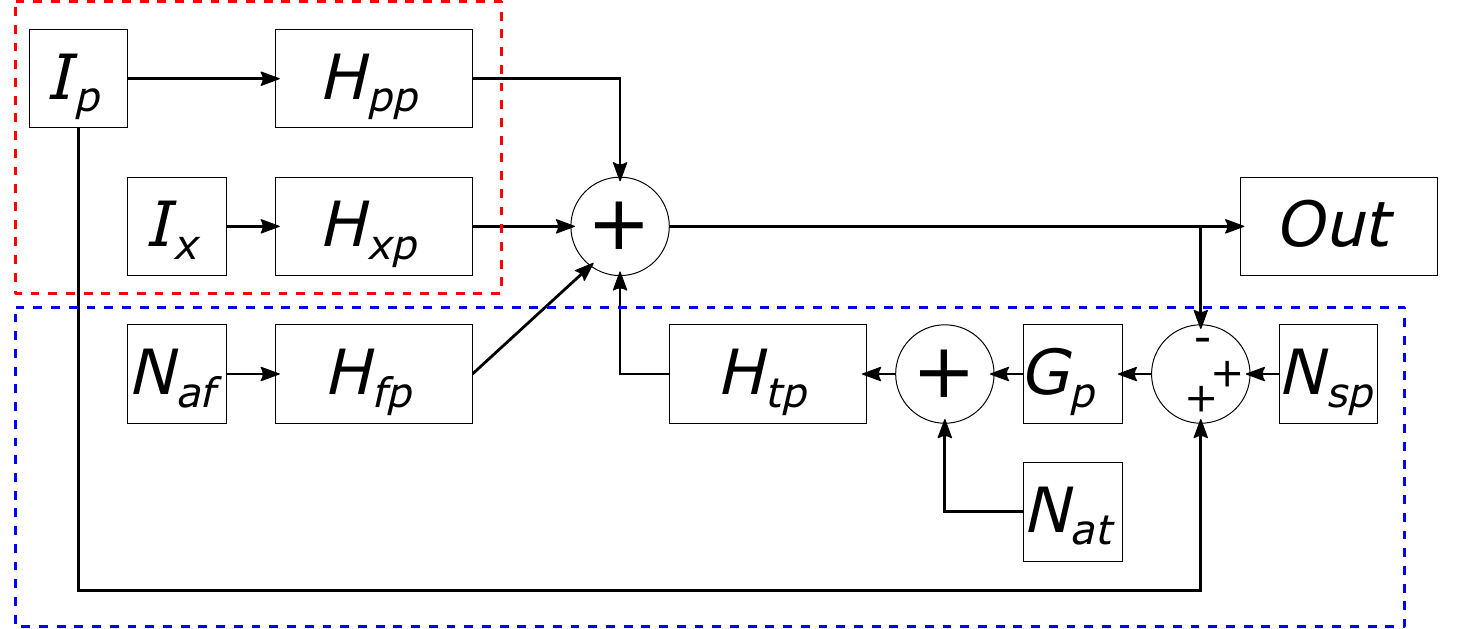}
	\caption{\label{fig_PitchNoiseModel}
		A schematic representation of the model used to calculate the noise budget in \textit{pitch}.}
\end{figure}

To estimate the noise performance and check compliance with the requirements during the design phase, we developed a noise model for each of the three actively controlled degrees of freedom (which are also the ones that most affect beam jitter and phase noise).
As an example, \cref{fig_PitchNoiseModel} shows a schematic of the noise model developed for the \textit{pitch} degree of freedom. The symbols are explained in \cref{tab_PNMsymbols}.

\begin{table}
	\caption{\label{tab_PNMsymbols}Symbols used in \cref{fig_PitchNoiseModel} and \cref{eq_PitchNoiseModel}.}
	\begin{tabular}{l p{0.7\linewidth}}
		Symbol & Explanation \\
		\hline 
		$H_{pp}$, $H_{xp}$	& transfer functions from \textit{x} and \textit{pitch} motion of the platform, respectively, to \textit{pitch} motion fo the optic\\ 
		$H_{fp}$, $H_{tp}$	& transfer functions from force and torque, respectively, to \textit{pitch} motion fo the optic \\
		$I_p$, $I_x$		& \textit{pitch} and \textit{x} motion of the platform \\ 
		$N_{af}$, $N_{at}$	& actuation force and torque noise, respectively \\
		$N_{sp}$			& \textit{pitch} sensing noise\\
		$G_p$				& feedback loop gain
	\end{tabular}
\end{table}

The top left part of the diagram represents the direct effect of motion of the suspension structure, in both \textit{pitch} and \textit{x}, on optic \textit{pitch} motion. The lower part represents the contribution of the AOSEMs: the differential \textit{pitch} readout between optic and suspension structure, affected by sensing noise, is conditioned by the feedback loop gain to obtain the \textit{pitch} torque actuation signal. This, together with torque and force actuation noise, affects the optic pitch through the relevant transfer functions. Although force and torque noise originating from the AOSEMs are not statistically independent, the correlation is small and we treated them as such for simplicity.

\begin{figure}
	\centering
	\includegraphics[width=\linewidth]{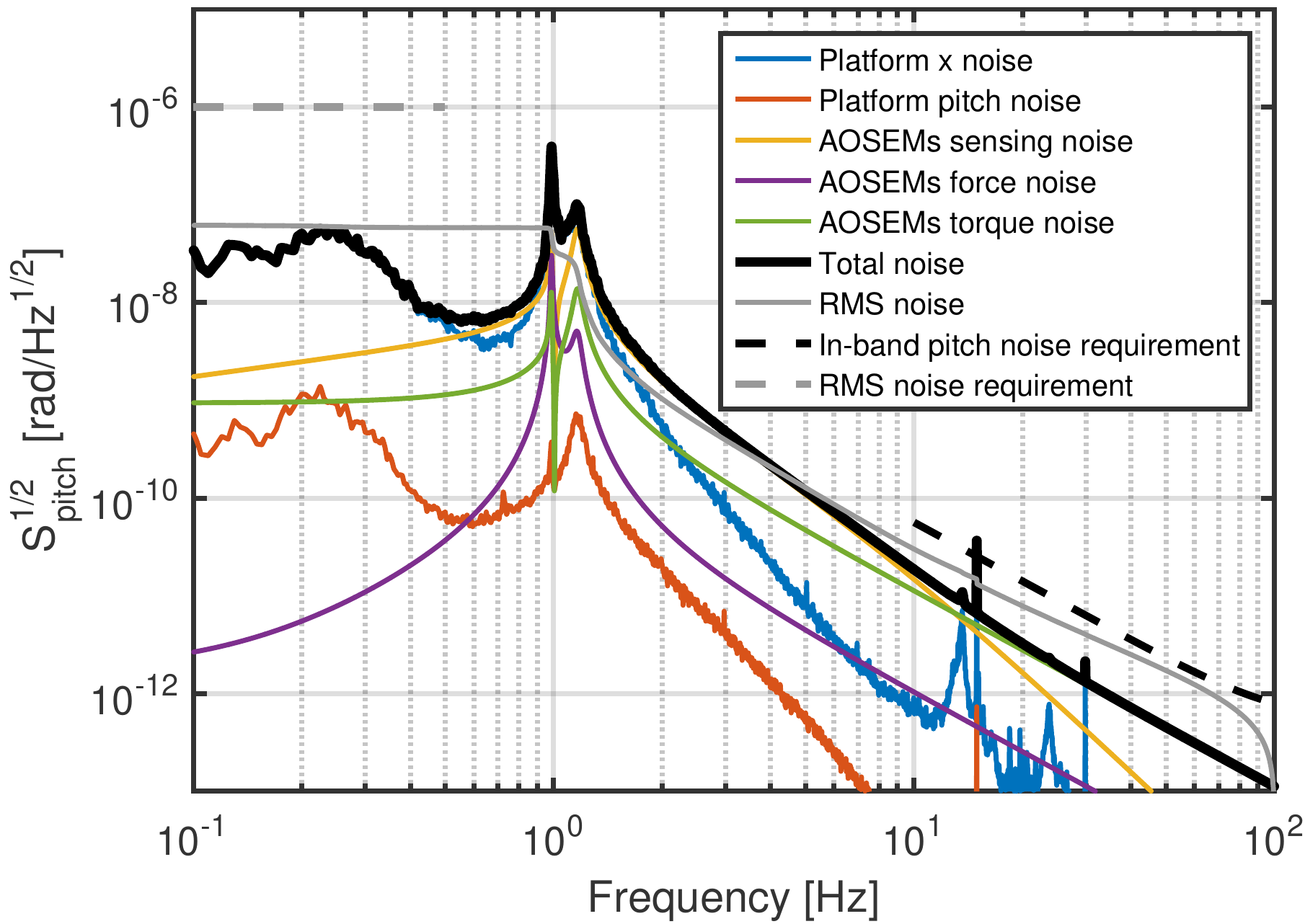}
	\caption{\label{fig_PitchNoiseBudget}
		(Color online) The noise budget of the \textit{pitch} degree of freedom based on the noise model represented in \cref{fig_PitchNoiseModel}. The black and grey dashed lines represent the requirements on the total pitch noise and its RMS value, respectively, in the bands of interest. For this example, we set $G_p$ equal to a simple one zero, two poles band-pass filter; it can be seen that even this simple solution is sufficient to damp the resonances and keep the RMS noise within specifications, without exceeding the in-band noise requirement.}
\end{figure}

Solving the model for the noise in \textit{pitch} yields:
\begin{multline}\label{eq_PitchNoiseModel}
N_{pitch} = \frac{1}{1+G_p H_{tp}} \Big( I_x H_{xp} + N_{af} H_{fp} + N_{at} H_{tp} + \\ 
+ N_{sp} G_p H_{tp} + I_p \left( H_{pp} + G_p H_{tp}\right)\Big)
\end{multline}
The various contributions to the final pitch noise, computed from input noise measured separately and transfer functions calculated from the Mathematica model, are plotted in \cref{fig_PitchNoiseBudget}. Here $G_p$ is a double-pole low pass filter; although this filter is not intended to be the final filter used in Advanced LIGO, it demonstrates that even this simple design is sufficient to meet requirements.

There are no specific requirements for the level of damping provided by the eddy current dampers; however, for non-cavity suspensions in Advanced LIGO it is generally considered adequate to obtain a quality factor, \textit{Q}, somewhere between \num{10} and \num{100} to prevent the amplitude of motion of the uncontrolled degrees of freedom to grow too big. We designed the eddy current dampers to be adjustable, so that their effect could be varied over a wide range of values. \cref{tab_ECDperf} shows the reduction of \textit{Q} in the three relevant degrees of freedom, measured by observing the rate of decay of oscillations, for the nominal placement of the magnets, \SI{2}{\mm} away from the optic holder.
As already mentioned at the end of \cref{sec_MecDesign}, the eddy current dampers also damp the actively controlled degrees of freedom, mostly \textit{x} and \textit{pitch}. This is not a problem in Advanced LIGO, because the resulting \textit{Q} is still higher than the value targeted by the active control.

\begin{table}
	\caption{\label{tab_ECDperf} Resonance quality factors measured with and without the eddy current dampers in their nominal position.}
	\begin{tabular}{l S S}
		\hline
		{DoF} & {Q, undamped} & {Q, damped} \\
		\hline 
		side/roll 1	& 6000\pm1000	& 74\pm2 \\ 
		side/roll 2	& 500\pm25		& 33\pm3 \\ 
		bounce		& 420\pm20		& 43\pm3 \\
		\hline 
	\end{tabular}
\end{table}

\subsection{Structural resonances}

\begin{figure}
	\includegraphics[width=\linewidth]{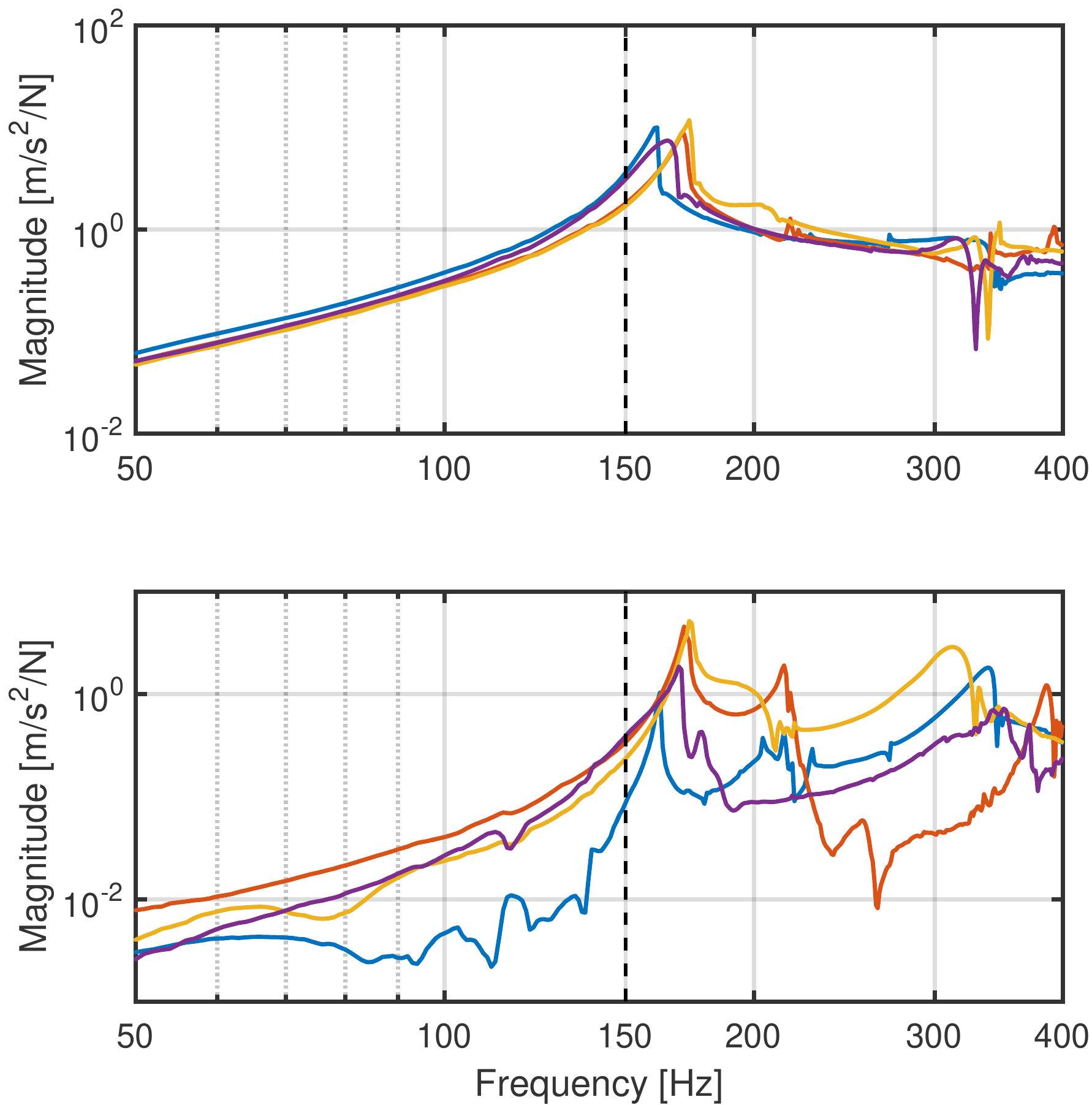}
	\caption{\label{fig_BKhammer}
		(Color online) Measurement of the mechanical resonances of the HAUX structure for all 4 units installed at the Hanford Observatory. A tri-axial accelerometer is attached to the structure, which is excited along the \textit{x} (top plot) and \textit{side} (bottom plot) directions using an calibrated hammer. In each plot, the four traces represent the four units tested at LHO. Each trace is the root mean square of the accelerations measured by the accelerometer along each of the three axis, so as to highlight any resonant peak independently of the specific shape of the mechanical mode excited.}
\end{figure}

The mechanical resonances of the structure depend on the rigidity of the structure itself, and on to what and how the structure is clamped. Using a commercial system from Br\"{u}el\&Kj\ae r,
we have measured each set of HAUX structural resonances when they are installed and clamped in their final position on the Advanced LIGO optical tables\cite{LHOBKhammer,LLOBKhammer}. As an example, \cref{fig_BKhammer} shows the measurements taken for all 4 units installed at the Hanford Observatory. We found the measurements to be very consistent among different suspensions and to meet the requirement that the lower resonance be above 150\,Hz, with the only exception of a unit installed at the Livingston Observatory; this non-compliance, probably due to a manufacturing issue, does not appear to have any significant impact on the performance of the ISI platforms, but it is nevertheless scheduled to be further investigated when interferometer operations allow.

\section{Conclusions}\label{sec:conclusions}
We have presented the design rationale and implementation of a compact single stage suspension for \SI{75}{\mm} diameter optics. The suspension provides isolation in all degrees of freedom, with resonant frequencies around \SI{1}{Hz} for all but the bounce and roll modes, which are below \SI{10}{Hz}. The suspension has active control for the three most critical degrees of freedom of the optic, and passive damping for the remaining three. The design incorporates a number of expedients to make installation/replacement and initial alignment of the optics more convenient. The suspension design meets the requirements for suspending out-of-cavity optics in the Advanced LIGO Input Optics chain, and is versatile enough to be used in other precision optic experiments.

\begin{acknowledgments}
The work presented in this paper has been supported by NSF grants PHY-0855313 and PHY-0969935, and by a subcontract stipulated with the LIGO Laboratory.
\end{acknowledgments}


%
%

%

\bibliography{HAUXbiblio}

\end{document}